\documentclass[
    ,final            
  ]
  {aipproc}

\layoutstyle{8x11single}

\usepackage{graphicx}

\newcommand{\beq}{\begin{eqnarray}}
\newcommand{\eeq}{\end{eqnarray}}
\newcommand{\bqa}{\begin{eqnarray}}
\newcommand{\eqa}{\end{eqnarray}}
\newcommand{\gsim}{\hspace*{0.2em}\raisebox{0.5ex}{$>$}
     \hspace{-0.8em}\raisebox{-0.3em}{$\sim$}\hspace*{0.2em}}

\begin{document}

\title{Universality in few-body systems with large scattering length}



\author{H.-W. Hammer}{
  address={Institute for Nuclear Theory, University of Washington, 
  Seattle, WA 98195-1550, USA}
}

\begin{abstract}
Effective Field Theory (EFT) provides a powerful framework that exploits
a separation of scales in physical systems to perform systematically
improvable, model-independent calculations.
Particularly interesting are few-body systems with
short-range interactions and large two-body scattering length.
Such systems display remarkable universal features. In systems with more
than two particles, a three-body force with limit cycle behavior is
required for consistent renormalization already at leading order.
We will review this EFT and some of its applications in the physics
of cold atoms and nuclear physics.
In particular, we will discuss the possibility of an infrared limit
cycle in QCD. Recent extensions of the EFT approach to the four-body
system and $N$-boson droplets in two spatial dimensions will also be
addressed.
\end{abstract}

\maketitle


\section{INTRODUCTION}

A separation of scales in a physical system can be exploited 
using the framework of Effective Field Theory (EFT) 
\cite{Kaplan:1995uv}. In EFT,
only low-energy (or long-range) degrees of freedom are included
explicitly, while all others are parametrized in terms of the most general
local contact interactions. This approach relies on the fact that
a low-energy probe of momentum $k$ cannot resolve structures on 
scales smaller than $1/k$. (Note that $\hbar=c=1$ in this talk.)
As a consequence, the influence of short-distance physics on 
low-energy observables can be captured in a few low-energy constants.
The EFT describes universal low-energy physics independent of
detailed assumptions about the short-distance dynamics. 

All low-energy observables can be calculated in a controlled expansion in 
powers of $kl$, where $l$ is the characteristic low-energy 
length scale of the system. Error estimates can be obtained from 
the higher orders in the expansion in $kl$.
The size of $l$ depends on the system 
under consideration: for particles interacting via
a finite range potential, e.g., $l$ is given by the range of the potential.
For the examples discussed in this talk, $l$ is of the order of 
the effective range $r_e$.

We will focus on applications of EFT to few-body systems
with large S-wave scattering length $|a| \gg l$. 
For a generic system, the scattering length
is of the same order of magnitude as the low-energy length scale $l$.
Only a very specific choice of the parameters in the underlying theory 
(a so-called {\it fine tuning}) will generate a large scattering length $a$.
Nevertheless, systems with large $a$ can be found in many
areas of physics. The fine tuning can be accidental or it can be 
controlled by an external parameter. 

\begin{figure}[ht]
\bigskip
\centerline{\includegraphics*[width=7cm,angle=0]{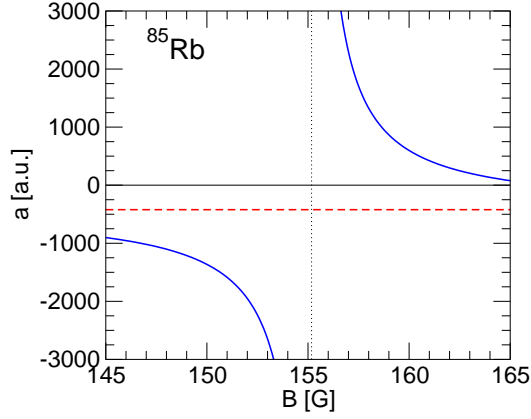}}
\caption{The scattering length $a$ of $^{85}$Rb atoms (in atomic units)
as a function of the magnetic field close to the Feshbach resonance at 
$B=155$ G (solid line). The dashed line indicates the off-resonant value
of $a$, while the vertical dotted line gives the resonance position.}
\label{fig:feshb}
\end{figure}

Examples with an accidental fine tuning are the S-wave scattering 
of nucleons and of $^4$He atoms. 
For S-wave nucleon-nucleon scattering,
the scattering lengths and effective ranges are
$a_s = -23.76$ fm and $r_s=2.75$ fm  in the spin-singlet channel
and $a_t = 5.42$ fm and $r_t=1.76$ fm in the spin-triplet channel.
The effective ranges are both comparable to the
characteristic low-energy length scale  $\ell_\pi =1/m_\pi\approx 1.4$ fm
which is given by the inverse pion mass.
However the scattering length $a_s$ is much larger than $\ell_\pi$
and $a_t$ is at least  significantly larger.
For $^4$He atoms, the scattering length $a\approx 104$\AA\
is more than a factor 10 larger than the typical low-energy length
scale $l\approx 5$\AA\ given by the 
van der Waals interaction. The scattering length of alkali atoms close 
to a Feshbach resonance can be tuned experimentally by adjusting the
external magnetic field. In Fig.~\ref{fig:feshb}, we show
the scattering length of $^{85}$Rb atoms as a function of the magnetic field
near the Feshbach resonance at $B=155$ G. Sufficiently close to the 
resonance, the scattering length can be made arbitrarily large.

At very low energies all these systems behave similarly and
show universal properties associated with the large 
scattering length.
A simple example in the two-body sector for $a>0$ is the existence
of a shallow molecule (the {\em dimer}) with binding energy
$B_2 = 1/(ma^2)\,$.
A particularly remarkable example in the three-body sector
is the existence of shallow three-body bound
states ({\it Efimov states}) with universal properties \cite{Efi71}.
Efimov states can cause dramatic dependence
of scattering observables on $a$ and on the energy \cite{Efi79,BrH04}.

We are interested in processes with typical momenta $k\sim 1/|a|$.
In this case, the EFT expansion becomes an expansion in powers of $l/|a|$. 
In leading order, we can simply set $l=0$ and three-body observables
are determined by $a$ and by a three-body parameter
that also determines the Efimov spectrum.
The higher-order corrections in $l/|a|$ can be calculated but will not 
be discussed in detail in this talk.

In the next section, we will briefly review the EFT for 
three-body systems with large scattering length $a$. 
In the following sections,
we will discuss some applications in nuclear and atomic physics. 
In particular, we will discuss the possibility of an infrared limit
cycle in QCD. Recent extensions of the EFT approach to the four-body
system and $N$-boson droplets in two spatial dimensions will also be
addressed. For a more detailed review, see  Ref.~\cite{BrH04}.

\section{THREE-BODY SYSTEM WITH LARGE SCATTERING LENGTH}

\begin{figure}[ht]
\bigskip
\centerline{\includegraphics*[width=12cm,angle=0]{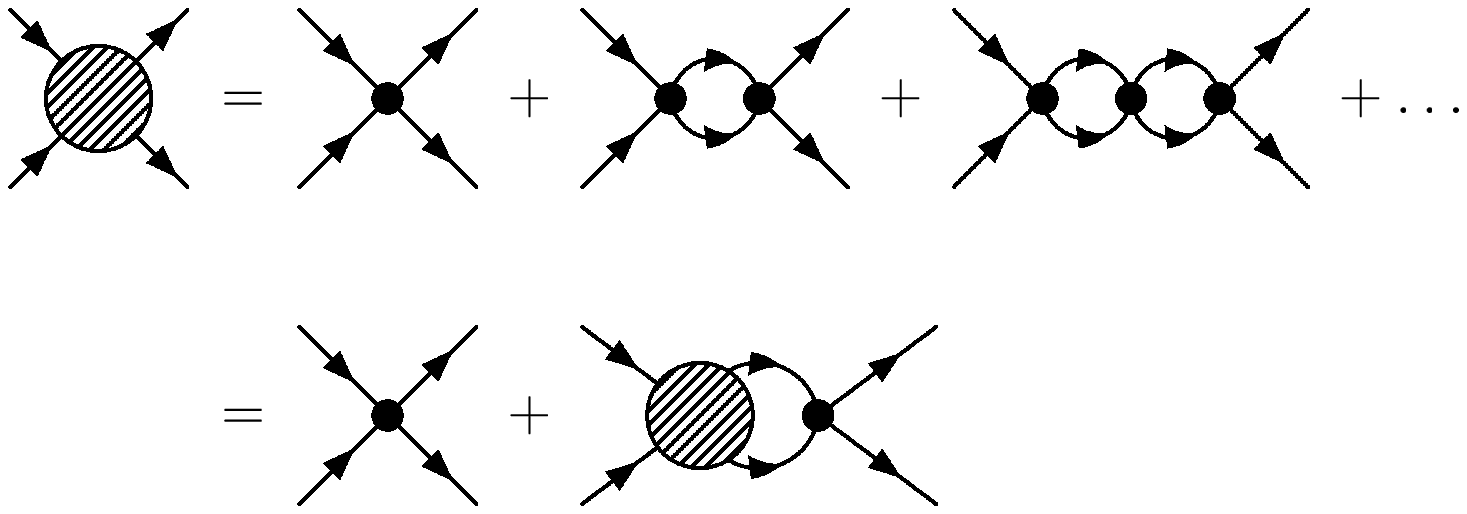}}
\caption{The bubble diagrams with the
$g_2$ interaction contributing to the two-body scattering amplitude.}
\label{fig:bubble}
\end{figure}

We start with with a two-body system of nonrelativistic bosons 
with large S-wave scattering length $a$ and mass $m$. 
The generalization to fermions is straightforward but will not be
discussed in detail.
In the following, we will refer to the bosons simply as atoms.
At sufficiently low energies, the most general Lagrangian 
may be written as:
\begin{equation}
{\cal L}  =  \psi^\dagger
             \bigg(i\partial_{t}+\frac{\vec{\nabla}^{2}}{2m}\bigg)\psi
 - \frac{g_2}{4} (\psi^\dagger \psi)^2
 - \frac{g_3}{36} (\psi^\dagger\psi)^3 + \ldots\,,
\label{eq:eftlag}
\end{equation}
where the dots represent higher-order derivative terms
which are suppressed at low energies.
The strength of the two-body interaction $g_2$ is determined by
the scattering length $a$, while the three-body interaction $g_3$ depends 
on a parameter to be introduced below. For momenta $k$ of the order of
the inverse scattering length $1/|a|$, the problem is nonperturbative 
in $ka$. The full two-atom scattering amplitude can be obtained
analytically by summing the so-called {\it bubble diagrams} with the
$g_2$ interaction term shown in Fig.~\ref{fig:bubble}.
The $g_3$ term does not contribute to two-body observables.
The bubble diagrams are ultraviolet divergent and have to be regulated.
This can be done by introducing an ultraviolet cutoff $\Lambda$ in the
loop integrals. The atom-atom scattering amplitude $f_{AA}$ can be renormalized
by allowing  the coupling $g_2$ to depend on $\Lambda$ and choosing 
$g_2(\Lambda)$ such that $f_{AA}$  is independent of the regulator.
After renormalization, the resulting amplitude reproduces the leading 
order of the well-known effective range expansion for the
scattering amplitude:~$f_{AA}(k)=(-1/a -ik)^{-1}\,,$
where the total energy is $E=k^2/m$
and $k$ is the center-of-mass momentum. 
If $a>0$, $f_{AA}$ has a pole at $k=i/a$ corresponding
to a shallow dimer with binding energy $B_2=1/(ma^2)$. 
Higher-order derivative
interactions are perturbative and give the momentum-dependent terms
in the effective range expansion. 

\begin{figure}[ht]
\bigskip
\centerline{\includegraphics*[width=10cm,angle=0]{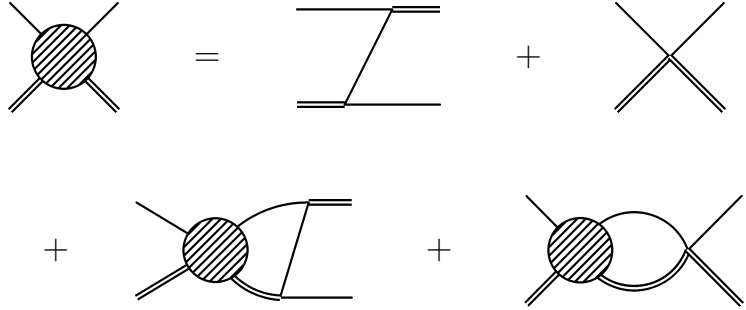}}
\caption{The integral equation for the atom-dimer scattering amplitude.
The single (double) line indicates the single-atom (two-atom)
propagator.}
\label{fig:ineq}
\end{figure}

We now turn to the three-body system. At leading order in $l/|a|$, 
the atom-dimer scattering amplitude is given by the 
integral equation shown in  Fig.~\ref{fig:ineq}. 
A solid line indicates the single-atom propagator and a double line 
indicates the full two-body propagator including all atom bubbles
(cf.~Fig.~\ref{fig:bubble}).
The integral equation contains contributions from both
the two-body and the three-body interaction terms $g_2$ and $g_3$, 
respectively.
The inhomogeneous term is given by the first two diagrams on the 
right-hand side: the one-atom exchange diagram
and the contribution of the three-body interaction. 
The integral equation simply sums these diagrams to all orders.
After projecting onto S-waves, we obtain the equation
\begin{eqnarray}
{\cal T} (k, p; E) & = & {16 \over 3 a}\, M(k,p;E)
+ {4 \over \pi} \int_0^\Lambda 
{dq \, q^2 \, M(q,p;E)\, {\cal T} (k, q; E)\over  -{1/a} + \sqrt{3q^2/4 -mE
-i \epsilon}}
\,,
\label{eq:BHvK}
\end{eqnarray}
for the off-shell atom-dimer scattering amplitude with the inhomogeneous
term 
\begin{eqnarray}
M(k,p;E)&=& {1 \over 2pk} \ln \left({p^2 + pk + k^2 -mE \over
p^2 - pk + k^2 - mE}\right) + {H(\Lambda) \over \Lambda^2}\,.
\label{eq:MkpE}
\end{eqnarray}
The logarithmic term is the S-wave projected one-atom exchange,
while the term proportional to $H(\Lambda)$ comes from the three-body
interaction. Note that we have introduced a cutoff regulator $\Lambda$ 
in Eq.~(\ref{eq:BHvK}). This cutoff is required to insure that 
(\ref{eq:BHvK}) has a unique solution. It is convenient to express $g_3$ as
\beq
g_3=-\frac{9g_2^2}{\Lambda^2}H(\Lambda)\,,
\eeq
which defines the function $H(\Lambda)$ in Eq.~(\ref{eq:MkpE}).
The S-wave atom-dimer scattering amplitude 
$f_{AD}$ is given by the 
solution ${\cal T}$ of  Eq.~(\ref{eq:BHvK}) evaluated at the on-shell 
point:
\beq
f_{AD}(k) = \frac{1}{k\cot\delta_0-ik}={\cal T} (k, k; E)\,, \qquad 
\mbox{where}\qquad E= \frac{3k^2}{4m}-\frac{1}{ma^2}
\eeq
is the total energy and $k$ is the center-of-mass momentum.
The three-body binding energies $B_3$ are given by those values of $E<0$
for which the homogeneous version of Eq.~(\ref{eq:BHvK}) has a
nontrivial solution.

All physical observables must be independent of the regulator
$\Lambda$. Therefore, we can determine $H(\Lambda)$ by varying 
$\Lambda$ and demanding invariance of the low-energy three-body observables
under this renormalization group transformation. This leads to the 
expression \cite{BHvK99,BHvK99b}:
\begin{eqnarray}
H (\Lambda) = {\cos [s_0 \ln (\Lambda/ \Lambda_*) + \arctan s_0]
\over \cos [s_0 \ln (\Lambda/ \Lambda_*) - \arctan s_0]}\,,
\label{H-Lambda}
\end{eqnarray}
where $s_0=1.00624...\,$ is a solution to the transcendental equation
\beq
s_0 \cosh\frac{\pi s_0}{2}=\frac{8}{\sqrt{3}}\sinh\frac{\pi s_0}{6}\,,
\eeq
and $\Lambda_*$
is a three-body parameter introduced by dimensional transmutation. 
This parameter cannot be predicted by the EFT and must be 
determined from a three-body observable.
\begin{figure}[ht]
\bigskip
\centerline{\includegraphics*[width=9cm,angle=0]{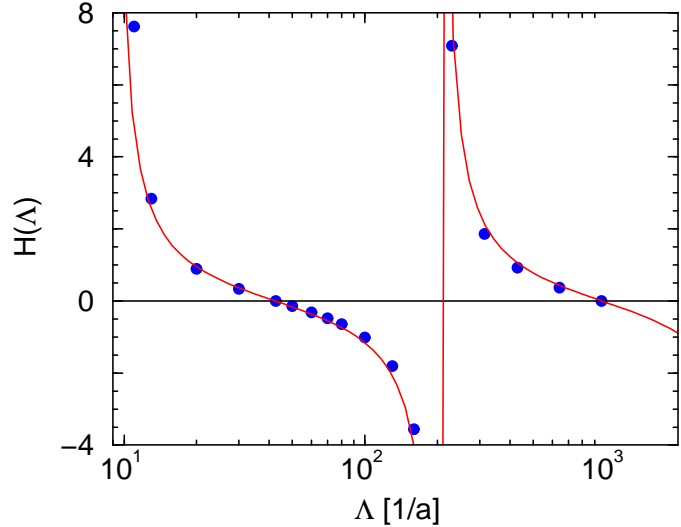}
}
\caption{
The three-body coupling $H$ as a function of the cutoff $\Lambda$
for a fixed value of the three-body parameter $\Lambda_*$.
The solid line shows the analytical expression (\ref{H-Lambda}), while
the dots show results from the numerical solution of Eq.~(\ref{eq:BHvK}).
}
\label{fig:limit}
\end{figure}
The dependence  of the three-body coupling $H$ on the cutoff $\Lambda$
is shown in Fig.~\ref{fig:limit}
for a fixed value of the three-body parameter $\Lambda_*$.
The solid line shows the analytical expression (\ref{H-Lambda}), while
the dots show results from the numerical solution of Eq.~(\ref{eq:BHvK}).
A good agreement between both methods is observed, indicating that the
renormalization of  Eq.~(\ref{eq:BHvK}) is well under control.
Note that $H (\Lambda)$ is periodic and runs on
a limit cycle. An important signature of an renormalization group (RG) 
limit cycle is discrete scale invariance.
When $\Lambda$ is increased by a factor of
$\exp(\pi/s_0)\approx 22.7$, $H (\Lambda)$ returns to its original 
value. Discrete scale invariance also arises in other contexts 
as varied as turbulence, sandpiles, earthquakes, 
and financial crashes \cite{Sor97}.
The possibility of renormalization group limit cycles
was first pointed out by Wilson in 1971 \cite{Wil71}, but no explicit
examples were known at the time. Recently, a number of field-theoretical
models with limit cycles have been constructed 
\cite{Glazek:2002hq,Glazek:2004,LeClair:2002ux,Leclair:2003xj}.
Leclair, Roman, and Sierra have used the colorful phrase 
{\it Russian doll renormalization group} to describe RG limit cycles
\cite{LeClair:2002ux,Leclair:2003xj}.
The name refers to a traditional souvenir from Russia consisting 
of a set of hollow wooden dolls that can be nested inside each other.
Similar to a limit cycle, the dolls show discrete scale invariance: the
scaling factors between each doll and the next smaller one are all 
approximately equal.

In summary, two parameters are required to specify a 
three-body system with large scattering length
at leading order in $l/|a|$:
they may be chosen as the scattering length $a$ (or equivalently
$B_2$ if $a>0$) and the three-body parameter $\Lambda_*$ 
\cite{BHvK99,BHvK99b}. 
In the following sections, we will discuss some applications of this
EFT in nuclear and atomic physics.

\section{UNIVERSAL PROPERTIES OF FEW-BODY SYSTEMS}

This EFT confirms and extends the universal predictions for
the three-body system first derived by Efimov \cite{Efi71,Efi79}. 
The best known example is
the {\it Efimov effect}, the accumulation of infinitely-many three-body 
bound states at threshold as $a\to\pm\infty$ \cite{Efi71}.
The ratio of the binding energies of successively shallower states 
rapidly approaches a constant $\lambda_0^2=\exp(2\pi/s_0)\approx 515$.  

However, universality also constrains three-body scattering observables.
The atom-dimer scattering length, e.g., can be expressed in terms of $a$
and $\Lambda_*$ as \cite{Efi79,BH02}
\begin{equation}
a_{12}=a\, (1.46-2.15\tan [s_0 \ln(a\Lambda_*) +0.09])\,
(1\,+\,{\cal O}(l/a))\,,\qquad\qquad a>0\,.
\end{equation}
Note that the log-periodic dependence of $H$ on $\Lambda_*$ 
is not an artefact of the renormalization and is also manifest in observables 
like $a_{12}$. This dependence could, e.g.,
be tested experimentally with atoms close to a Feshbach resonance
by varying $a$. Similar expressions can be obtained for other three-body
observables such as scattering phase shifts as well as rates for
three-body recombination and dimer relaxation 
(including the effects of deeply-bound two-body states) \cite{BH03}.

\begin{figure}[ht]
\centerline{\includegraphics*[width=9cm,angle=0]{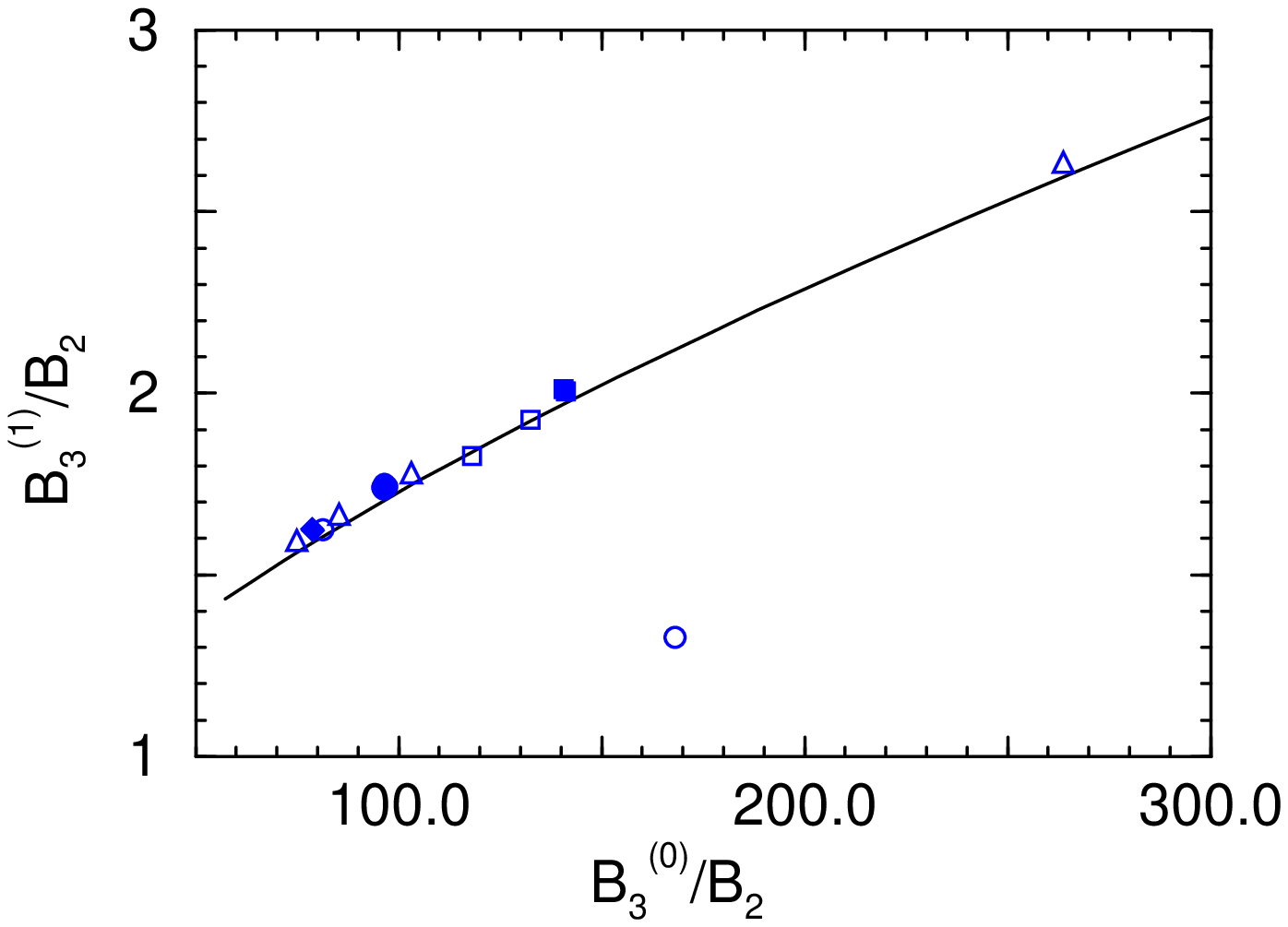}
\quad\includegraphics*[width=7.4cm,angle=0]{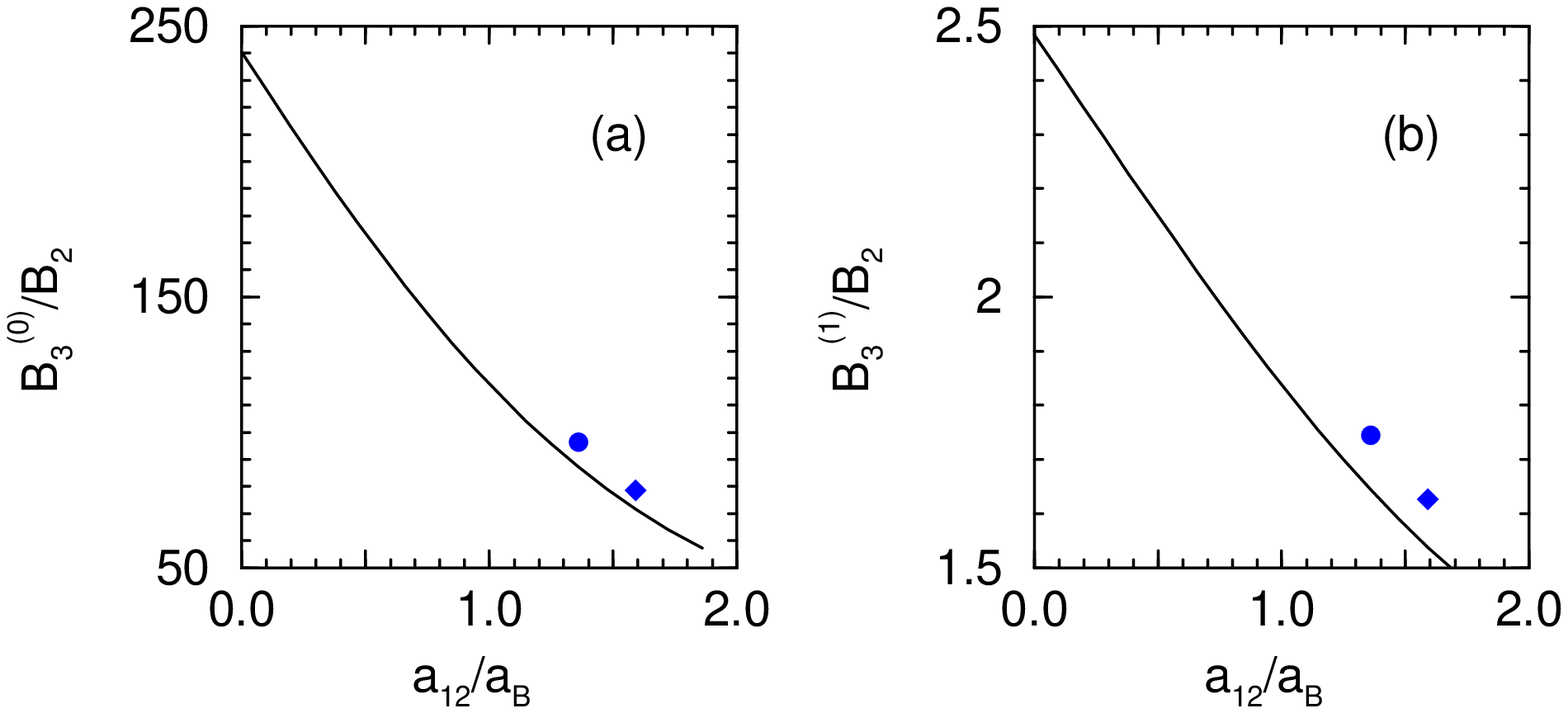}}
\caption{The scaling functions relating the $^4$He trimer ground and 
excited state energies (left panel) and the $^4$He trimer ground state 
and the atom-dimer scattering length (right panel).
The data points are calculations using various methods and $^4$He 
potentials (see Ref.~\cite{MSSK01} and references therein).
Note that $a_B\equiv 1/\sqrt{mB_2}$.
}
\label{fig:scale3}
\end{figure}

Since up to corrections of order $l/|a|$,
low-energy three-body observables depend on $a$ and $\Lambda_*$ only, 
they obey non-trivial scaling relations. If dimensionless combinations
of such observables are plotted against each other, they must fall close
to a line parametrized by $\Lambda_*$ \cite{BHvK99b,BH02}.
An example of a scaling function relating $^4$He trimer ground and excited 
state energies $B_3^{(0)}$ and $B_3^{(1)}$ is shown in the left panel
of Fig.~\ref{fig:scale3}. 
(A related scaling function was obtained in Ref.~\cite{FTD99}.)
The data points show calculations using various approaches and $^4$He 
potentials. Since different potentials have approximately the same
scattering length but include different short-distance physics,
different points on this line correspond to different values of $\Lambda_*$.
The small deviations of the potential model calculations are mainly due to
effective range effects. They are of the order $r_e/a \approx 10\%$
and can be calculated at next-to-leading order in EFT.
The calculation corresponding to the data point far off the universal curve 
can easily be identified as problematic since the deviation from 
universality
by far exceeds the expected 10\%.  The right panel shows
the scaling function relating the $^4$He trimer ground state 
energy $B_3^{(0)}$  and the atom-dimer scattering length $a_{12}$.
A similar scaling relation
is observed in nuclear physics between the spin-doublet neutron-deuteron
scattering length and the triton binding energy and is known
as the Phillips line \cite{Phillips68}.

\begin{figure}[ht]
\centerline{\includegraphics*[width=8cm,angle=0]{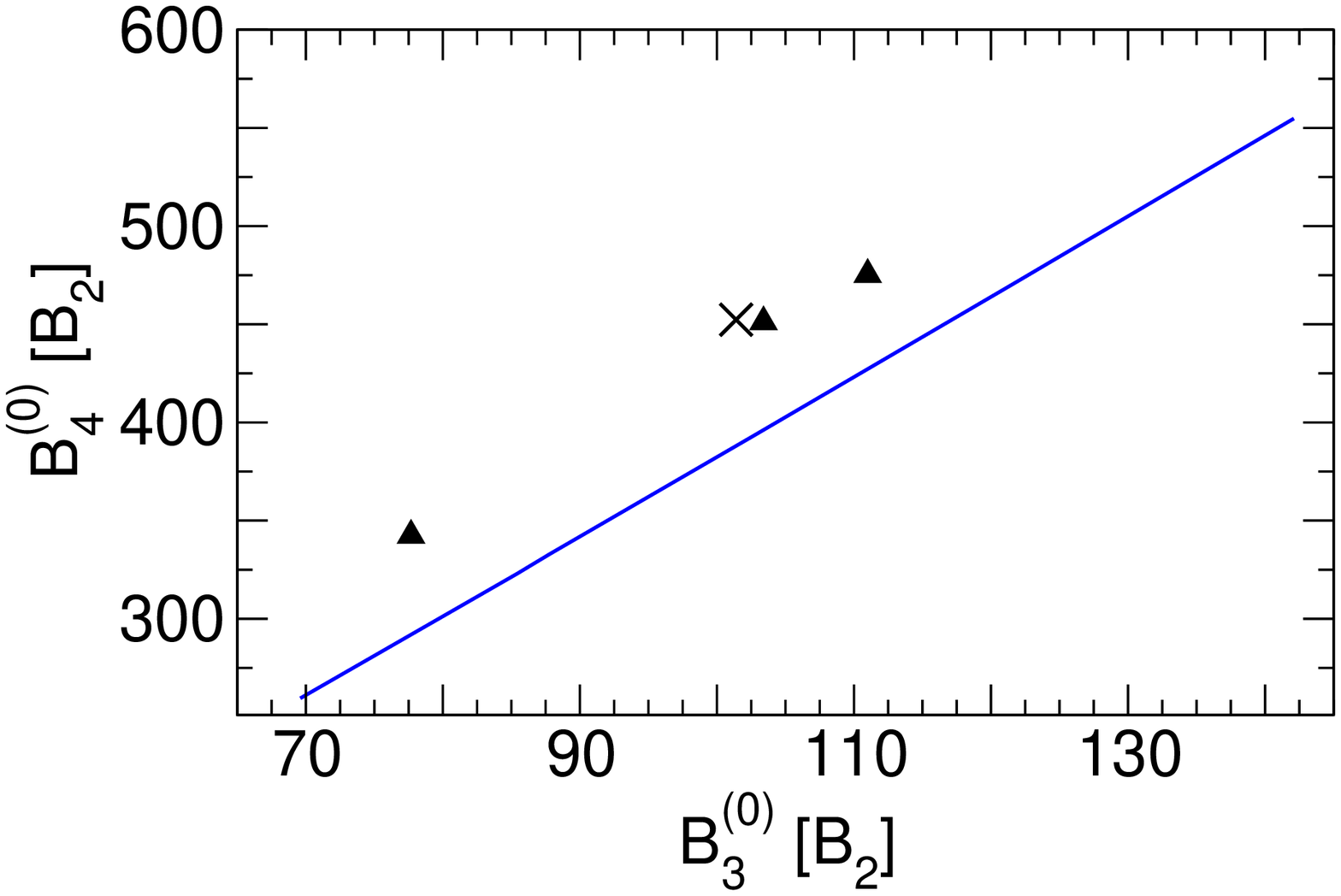}
\quad
\includegraphics*[width=8cm,angle=0]{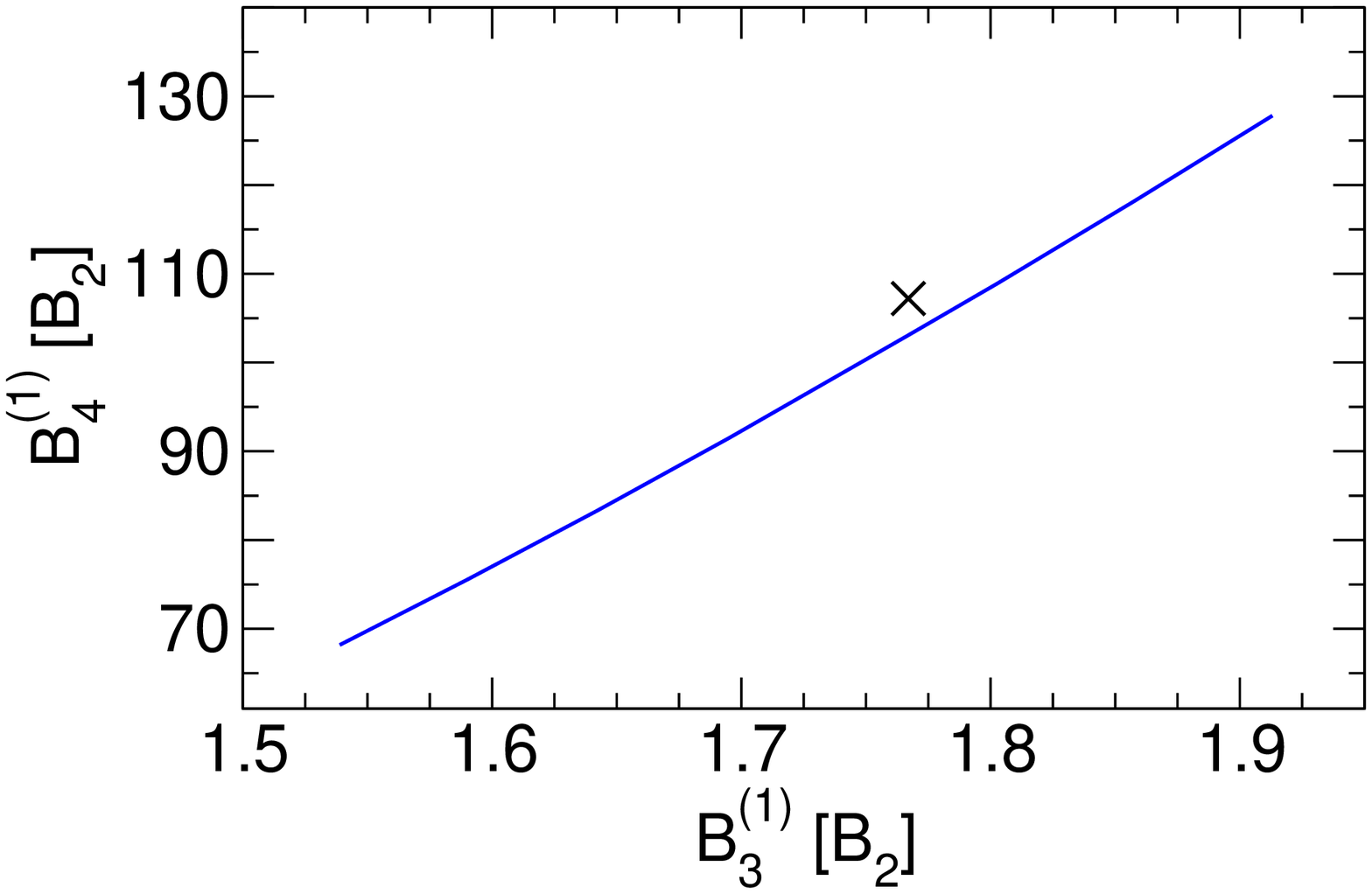}}
\caption{The scaling function relating the $^4$He trimer and tetramer 
ground state (left panel) and excited state (right panel)
energies. The data points are calculations using various methods and 
$^4$He potentials (see Ref.~\cite{Blume:2000} and references
therein).
}
\label{fig:scale4}
\end{figure}

Recently, we have extended the effective theory for large scattering 
length to the four-body system \cite{Platter:2004qn}. 
It is advantageous in this case to use an effective
quantum mechanics framework, since one can start directly from the 
well-known Yakubovsky equations. We have shown that no four-body
parameter enters at leading order in $l/|a|$. Therefore 
renormalization of the three-body system automatically guarantees 
renormalization of the four-body system. As a consequence, there
are universal scaling functions relating three- and four-body
observables as well. As an example, we show the 
correlation between the trimer and tetramer binding energies
in Fig.~\ref{fig:scale4}.
The left panel shows the correlation between the ground state
energies while the right panel shows the correlation between the 
excited state energies. The data points are calculations using various 
methods and $^4$He potentials (see Ref.~\cite{Blume:2000} and references
therein). We conclude that universality is well satisfied in the three-
and four-body systems of $^4$He atoms.

The existence of these scaling functions is a universal feature 
of systems with large scattering length and is independent of the 
details of the short-distance physics. Similar correlations
between few-body observables appear for example also in nuclear
systems. The correlation between the triton and $\alpha$ particle
binding energies (the Tjon line \cite{Tjo75}) 
and the Phillips line can be explained using a similar
effective theory \cite{Platter:tjon,EfiTk88,Bedaque:1999ve}.
The spin-singlet and spin-triplet scattering lengths 
$a_s$ and $a_t$ for nucleons
are both significantly larger than the range of the nuclear force. 
This observation can be used as the basis for an EFT approach
to the few-nucleon problem in which the nuclear forces are approximated
by contact interactions with strengths adjusted 
to reproduce the scattering lengths $a_s$ and $a_t$ 
\cite{Bedaque:1999ve}. 
(For an earlier related approach, see Ref.~\cite{Efi81}.)
This EFT does not contain explicit pion degrees
of freedom and is sometimes referred to as {\it pionless EFT}.
The EFT involves an isospin doublet $N$
of Pauli fields with two independent two-body contact interactions:
$N^\dagger \sigma_i N^c N^{c \dagger} \sigma_i N$ 
and  $N^\dagger \tau_k N^c N^{c \dagger} \tau_k N$,
where $N^c = \sigma_2 \tau_2 N^*$.
Renormalization in the two-body sector requires 
the two coupling constants to be adjusted as functions of 
$\Lambda$ to obtain the correct values of $a_s$ and $a_t$.
Renormalization in the three-body sector requires 
a three-body contact interaction 
$N^\dagger \sigma_i N^c N^{c \dagger} \sigma_j N 
	N^\dagger \sigma_i \sigma_j N$
with a coupling constant proportional to 
(\ref{H-Lambda}) \cite{Bedaque:1999ve}.
Thus the renormalization involves an ultraviolet limit cycle.
The scaling-violation parameter $\Lambda_*$ can be determined by
using the triton binding energy $B_t$ as input.
Effective range effects and other higher order corrections 
can be treated in perturbation theory \cite{Bedaque:2002yg}.

\begin{figure}[ht]
\centerline{\includegraphics*[width=10cm,angle=0]{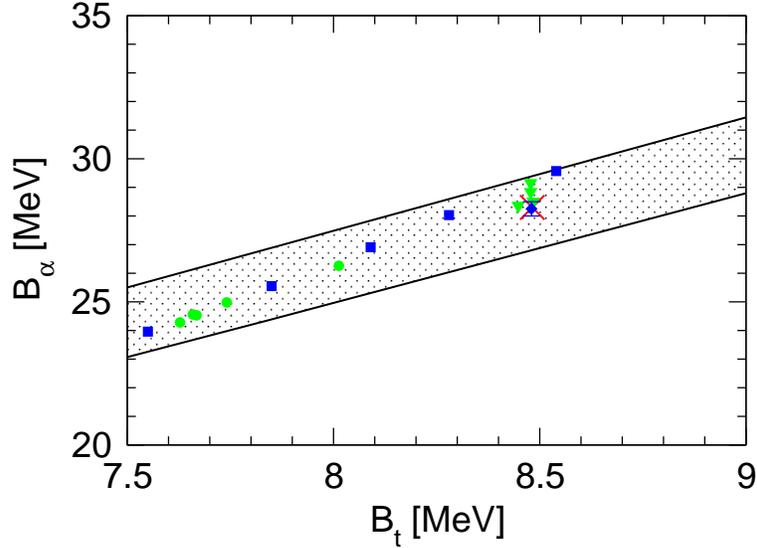}}
\caption{\label{fig:tjon}
The correlation between the binding energies of the 
triton and the $\alpha$ particle (the Tjon line). The lower (upper) line
shows our leading order result using  $a_s$ and $B_d$
($a_s$ and $a_t$) as two-body input. The grey circles and triangles show
various calculations using phenomenological potentials \cite{Nogga:2000uu}.
The squares show the results of chiral EFT at NLO for different cutoffs
while the diamond shows the N$^2$LO result 
\cite{Epelbaum:2000mx,Epelbaum:2002vt}.
The cross shows the experimental point.
}
\end{figure}

In Fig.~\ref{fig:tjon}, we show the result for the Tjon line with $a_s$
and $B_d$ as input (lower line)
and $a_s$ and $a_t$ as two-body input (upper line). Both lines generate
a band that gives a naive estimate of higher order corrections
in $\ell/|a|$. We also show some calculations using phenomenological 
potentials \cite{Nogga:2000uu} and a chiral EFT potential with explicit
pions \cite{Epelbaum:2000mx,Epelbaum:2002vt}.
All calculations with interactions that give a
large scattering length must lie close to this line. Different
short-distance physics and/or cutoff dependence should only move 
the results along the Tjon line. This can for example be observed in
the NLO results with the chiral potential indicated by the squares
in Fig.~\ref{fig:tjon} or in
the few-body calculations with the low-momentum $NN$ potential 
$V_{{\rm low-}k}$ carried out in Ref.~\cite{Nogga:2004ab}. The 
$V_{{\rm low-}k}$ potential
is obtained from phenomenological $NN$ interactions by intergrating out 
high-momentum modes above a cutoff $\Lambda$
but leaving two-body observables (such as the large
scattering lengths) unchanged. The results of few-body calculations 
with $V_{{\rm low-}k}$ are not independent of $\Lambda$ but lie all
close to the Tjon line (cf. Fig.~2 in Ref.~\cite{Nogga:2004ab}).

If we take the triton binding energy $B_t=8.48$ MeV as input, we can 
use the curves in Fig.~\ref{fig:tjon} to predict the binding energy
of the $\alpha$ particle. 
If the spin-singlet and spin-triplet scattering lengths $a_s$ and $a_t$
are used as the two-body input, we find
$B_\alpha = 29.5$~MeV. If the deuteron binding energy $B_d$
is used as input instead of $a_t$, 
we obtain $B_\alpha = 26.9$ MeV. This variation
is consistent with the expected 30\% accuracy of a leading
order calculation in $\ell/|a|$. Our results agree
with the (Coulomb corrected) experimental value
$B_\alpha^{exp}=(29.0 \pm 0.1)$~MeV to within 10\%. We conclude
that the ground state energy of the $\alpha$ particle can be described
by an effective theory with short-range interactions only.
Of course, a more refined analysis should also include Coulomb
corrections. However, the size of these corrections is expected
to be smaller than the NLO contribution from the two-body effective ranges.
Furthermore, corrections due to isospin violation will appear naturally
at higher order within the effective theory.

\section{AN INFRARED LIMIT CYCLE IN QCD}

\begin{figure}[ht]
\centerline{\includegraphics*[width=12cm,angle=0,clip=true]{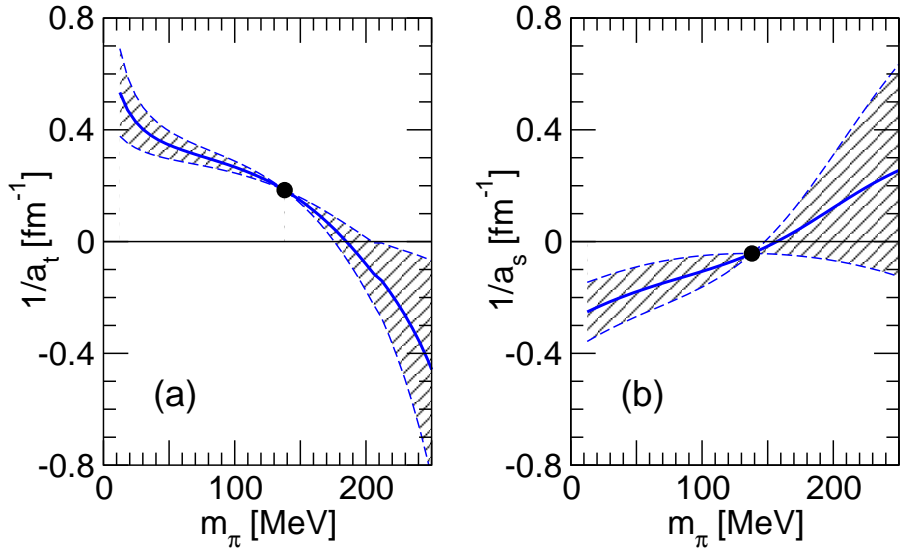}}
\caption{The inverse scattering lengths $1/a_t$ (left panel) and
$1/a_s$ (right panel) as functions of $m_\pi$ as predicted by the EFT
with pions of Ref.~\cite{Epelbaum:2002gb}.}
\label{fig:1oa}
\end{figure}

The low-energy few-nucleon problem can also be described 
by an EFT that includes explicit pion fields.
Such an EFT has been used to extrapolate nuclear forces 
to the chiral limit of QCD in which the pion is exactly 
massless \cite{Beane:2001bc,Beane:2002vs,Beane:2002xf,Epelbaum:2002gb}. 
In Fig.~\ref{fig:1oa}, we show the results for the extrapolation
of the inverse scattering lengths $1/a_t$ and $1/a_s$
as functions of $m_\pi$ from the calculation by
Epelbaum et al.~\cite{Epelbaum:2002gb}.
The extrapolation to larger values of $m_\pi$ predicts
that $a_t$ diverges and 
the deuteron becomes unbound at a critical 
value in the range 170 MeV $< \, m_\pi \, <$ 210 MeV.
It is also predicted that $a_s$ is likely to diverge
and the spin-singlet deuteron become bound 
at some critical value of $m_\pi$ not much larger than 150 MeV. 
Both critical values are close to the physical value
$m_\pi = 135$ MeV.  
Beane and Savage have argued that the errors in the extrapolation
to the chiral limit are larger than estimated in 
Ref.~\cite{Epelbaum:2002gb}, but their errors agree for the small 
extrapolation to the critical values \cite{Beane:2002vs,Beane:2002xf}.

\begin{figure}[ht]
\centerline{\includegraphics*[width=10cm,angle=0,clip=true]{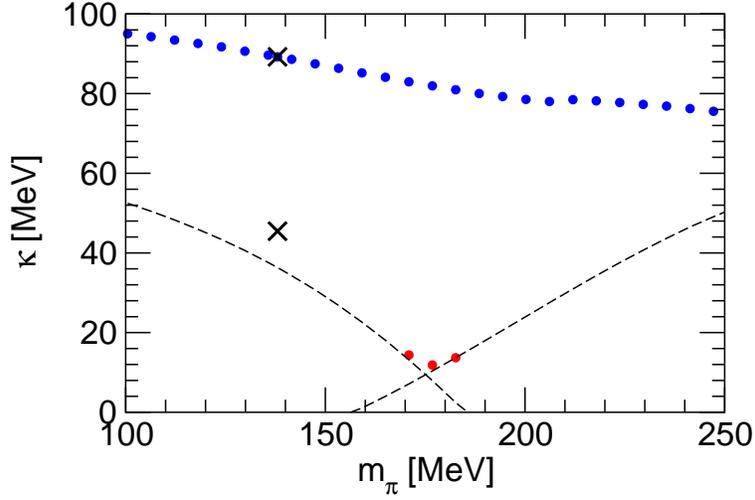}}
\caption{The binding momenta $\kappa=(mB_3)^{1/2}$ 
of $p n n$ bound states as a function of $m_\pi$.
The circles are the triton ground state and excited state. 
The crosses are the physical binding energies of the 
deuteron and triton.  The dashed lines are the thresholds
for decay into a nucleon plus a deuteron (left curve) or 
a spin-singlet deuteron (right curve).
}              
\label{fig:spec}
\end{figure}
Chiral extrapolations can also be calculated 
using the EFT without pions \cite{Braaten:2003eu}.
The inputs required are the chiral extrapolations $a_s(m_\pi)$, 
$a_t(m_\pi)$, and $B_t(m_\pi)$,
which can be calculated using an EFT with pions.
As an illustration, we take the central values of the error
bands for the inverse scattering lengths $1/a_s(m_\pi)$ and $1/a_t(m_\pi)$ 
from the chiral extrapolation in Ref.~\cite{Epelbaum:2002gb}.
Since the chiral extrapolation of the 
triton binding energy $B_t(m_\pi)$ has not yet been calculated and
since $\Lambda_*$ should vary smoothly with $m_\pi$,
we approximate it by its physical value $\Lambda_* =189$ MeV 
for $m_\pi=138$ MeV. In Fig.~\ref{fig:spec}, 
we show the resulting three-body spectrum in the triton 
channel as a function of $m_\pi$.  
Near $m_\pi \approx 175$ MeV where the decay threshold 
comes closest to $\kappa = 0$,
an excited state of the triton appears.
This excited state is a hint 
that the system is very close to an infrared limit cycle.
In the case illustrated by Fig.~\ref{fig:spec}, the value of $m_\pi$
at which $a_t$ diverges is larger than that at which $a_s$ diverges.
If they both diverged at the same value of $m_\pi$,
there would be an exact infrared limit cycle.

We conjecture that QCD can be tuned to this infrared limit cycle 
by adjusting the up and down quark masses 
$m_u$ and $m_d$ \cite{Braaten:2003eu}.
As illustrated in Fig.~\ref{fig:spec}, the tuning of $m_\pi$,
which corresponds to $m_u + m_d$, is likely to bring 
the system close enough to the infrared limit cycle for the triton
to have one excited state.  We conjecture that by adjusting
the two parameters $m_u$ and $m_d$ to critical values,
one can make $a_t$ and $a_s$ diverge simultaneously.
At this critical point, the deuteron and spin-singlet deuteron would both 
have zero binding energy and the Efimov effect \cite{Efi71} would 
occur: the triton would have 
infinitely-many increasingly-shallow excited states.
The ratio of the binding energies of successively shallower states 
would rapidly approach a constant $\lambda_0^2$ close to 515.  
The limit cycle would be manifest in the Efimov effect for the triton.
This argument requires that QCD with the appropriate combination of up
and down quark masses is in the same universality class as the
three-body system with large scattering length \cite{Wilson:2004de}.
It would be interesting to demonstrate the existence of this 
infrared RG limit cycle in QCD using Lattice QCD and EFT.

\section{$N$-BOSON DROPLETS IN TWO DIMENSIONS}

In this section, we consider the universal properties of
weakly interacting bosons with large scattering length
(or equivalently a shallow dimer state)
in two spatial dimensions (2D) \cite{Hammer:2004as}.
The 2D case is very different from the three-dimensional
case discussed above. In particular, there is no Efimov effect 
or Thomas collapse and no three-body interaction is
required at leading order. 
We consider self-bound droplets of $N(\gg1)$
bosons interacting weakly via an {\em attractive}, short-ranged 
pair potential. (This corresponds to Eq.~(\ref{eq:eftlag}) with
$g_3$ set to zero.)
Our analysis relies strongly on the
property of asymptotic freedom of 2D bosons with an attractive
interaction.

In 2D, any attractive potential has at least one bound state.  For the
potential $-g\delta^2(r)$ with small $g>0$, there is exactly one bound
state with an exponentially small binding energy,
\beq
  B_2 \sim \Lambda^2\exp\left( - {4\pi}/g\right),
\eeq
where $\Lambda$ is the ultraviolet momentum cutoff (which is the
inverse of the range of the potential).
Asymptotic freedom provides an elegant way to understand
this result.  In 2D nonrelativistic theory, the four-boson
interaction $g(\psi^\dagger\psi)^2$ is marginal.  The coupling runs
logarithmically with the length scale $R$, and the running can be
found by performing the standard renormalization group (RG) procedure.
For $g>0$, the coupling grows in the infrared, in a manner similar to 
the QCD coupling \cite{Gross:1973id,Politzer:fx}.
The dependence of the coupling on the length scale $R$ is given by
\begin{equation}\label{running}
  g(R) = \left[\frac{1}{g} - \frac{1}{4\pi}\ln 
        (\Lambda^2 R^2)\right]^{-1}\,,
\end{equation}
so the coupling becomes large when $R$ is comparable to the size of
the two body bound state $B_2^{-1/2}$. This is in essence the
phenomenon of dimensional transmutation: a dynamical scale is
generated by the coupling constant and the cutoff scale.
It is natural, then, that $B_2$ is the only physical energy scale in
the problem: the binding energy of three-particle, four-particle,...
bound states are proportional to $B_2$.  
(Remember that in 2D there is no Efimov effect
and no three-body parameter is required.) 
The $N$-particle binding energy $B_N$, however, can be very different 
from $B_2$ if $N$ is parametrically large.  We use the
variational method to calculate the size of the bound state.  For a
cluster of a large number of bosons, one can apply classical
field theory.  We thus have to minimize the expectation
value of the Hamiltonian with respect to all field configurations $\psi(r)$
satisfying the constraint $N = \int\!d^2r\, \psi^\dagger\psi\,$.
This is equivalent to a Hartree calculation with the running
coupling constant $g(R)$ instead of the bare one. In the limit
of a large number $N$ of particles in the droplet, some exact 
predictions can be obtained.

The system possesses surprising universal
properties.  Namely, if one denotes the size of the $N$-body droplet
as $R_N$, then at large $N$ and in the limit of zero range of the
interaction potential \cite{Hammer:2004as}:
\begin{equation}
  \label{RNratio}
  {R_{N+1}\over R_N} \approx 0.3417,\qquad 
  {B_{N+1}\over B_N} \approx 8.567, \qquad 
  N\gg 1\,.
\end{equation}
The size of the bound state decreases exponentially with $N$: adding a
boson into an existing $N$-boson droplet reduces the size of the
droplet by almost a factor of three.  Correspondingly, the binding energy of
$N$ bosons $B_N$ increases exponentially with $N$.
This implies that the energy required to remove
one particle from a $N$-body bound state (the analog of the
nucleon separation energy for nuclei) is about 88\% of the
total binding energy.  This is in contrast to most other physical
systems, where separating one particle costs much less energy than the
total binding energy, provided the number of particles in the bound
state is large. Equations (\ref{RNratio}) are accurate to leading order
in $1/N$ but the $1/N$-corrections are calculable.

For the universal predictions (\ref{RNratio}) to apply
in realistic systems with finite-range interactions,
the $N$-body bound states need to be sufficiently shallow and hence 
have a size $R_N$ large compared to the natural low-energy length scale $l$.
Depending on the physical system,
$l$ can be the van der Waals length, the range of the 
potential or some other scale. As a consequence,
Eqs.~(\ref{RNratio}) are valid in realistic systems
for $N$ large, but below a critical value,
\begin{equation}
  1 \ll N \ll N_{\rm crit} \approx 0.931 \ln({R_2}/{l})
  + {\cal O}(1)\,.
\label{eq:break}
\end{equation}
At $N=N_{\rm crit}$ the size of the droplet is comparable to $l$
and universality is lost.  If there is an exponentially
large separation between $R_2$ 
and $l$, then $N_{\rm crit}$ is much larger than one and 
the condition (\ref{eq:break}) can be satisfied.

\begin{figure}[ht]
\centerline{\includegraphics*[width=10cm,angle=0,clip=true]{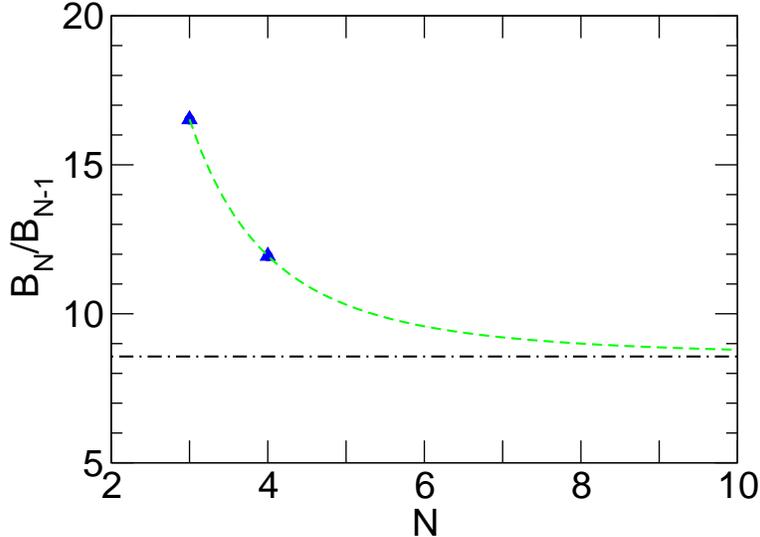}}
\caption{The ratio $B_N/B_{N-1}$ as a function of $N$. The dot-dashed line 
shows the asymptotic value of $8.567$. The dashed line is an estimate of how
this value is approached as $N$ increases.}
\label{fig:bind}
\end{figure}

We can compare our prediction with exact few-body calculations
for $N=3,4$. While the $1/N$-corrections are expected to be
relatively large in this case, we can estimate how the universal result 
for ${B_{N+1}}/{B_N}$  is approached.
The three-body system for a zero-range potential in 2D has exactly two
bound states: the ground state with
$B_3^{(0)}=16.522688(1)\, B_2$ and one excited state with
$B_3^{(1)}=1.2704091(1)\, B_2$ \cite{Bruch,Nielsen,Hammer:2004as}.
Similarly, the four-body system for a zero-range potential in 2D 
has two bound states: the ground 
state with $B_4^{(0)}=197.3(1)\,B_2$ and one excited state with 
$B_4^{(1)}=25.5(1)\,B_2$ \cite{Platter:2004ns}.
The prediction (\ref{RNratio}) applies to the ground state energies
$B_3^{(0)}$ and $B_4^{(0)}$. The exact few-body results for $N=3,4$
are compared to the asymptotic prediction for $B_N/B_{N-1}$
indicated by the dot-dashed line in Fig.~\ref{fig:bind}.
The ratio $B_3^{(0)}/B_2\approx 16.5$ is almost twice as large as the 
asymptotic value~(\ref{RNratio}), while the ratio $B_3^{(0)}/B_4^{(0)}
\approx 11.9$ is already considerably closer.
These deviations are expected for such small 
values of $N$. Note, however, that the ratio of
the root mean square radii of the two- and three-body wave functions
is $0.306$~\cite{Nielsen}, close to the asymptotic
value~(\ref{RNratio}). 
The dashed line gives an estimate of how the
large-$N$ value should be approached \cite{Platter:2004ns}. 
This estimate assumes the expansion:
\beq
B_N = \left(c_0 +\frac{c_{-1}}{N}+\frac{c_{-2}}{N^2}
+\ldots\right)\, 8.567^N \,,
\label{eq:BNexp}
\eeq
leading to
\beq
\frac{B_N}{B_{N-1}}= 8.567 +{\cal O}\left(N^{-2}\right)\,.
\label{eq:BNratioexp}
\eeq
The dashed line in  Fig.~\ref{fig:bind} was obtained by fitting the
coefficients of the $1/N^2$ and $1/N^3$ terms in Eq.~(\ref{eq:BNratioexp})
to the data points for $N=3,4$.

It would be interesting to test the universal 
predictions (\ref{RNratio}) both theoretically and 
experimentally for $N>4$. On the theoretical side, Monte Carlo 
techniques appear to be a promising avenue.
Furthermore, the experimental realizablity of self-bound 
2D boson systems with weak interactions should be investigated. 
According to the estimate in Fig.~\ref{fig:bind}, the 
$1/N$ corrections to Eqs.~(\ref{RNratio}) are small
for $N\gsim 6$. Using (\ref{eq:break}),
this requires $R_2/\ell \gg 600$. We are not aware of
any physical system that satisfies this constraint. However, such
a system could possibly be realized close to a Feshbach resonance
where $R_2$ can be made arbitrarily large.

\section{SUMMARY}

We have discussed the EFT for few-body systems with
short-range interactions and large scattering length $a$. 
The renormalization of the three-body system with large $a$
in three spatial dimensions
requires a one-parameter three-body force governed by a limit
cycle already at leading order in the expansion in $l/|a|$.
As a consequence, two parameters are required to specify a three-body 
system: the scattering length $a$ (or the dimer
binding energy $B_2$) and the three-body parameter $\Lambda_*$.
Once these two parameters are given, the properties of the 
three- and four-body systems are fully 
determined at leading order in $l/|a|$.

The large scattering length leads to universal properties
independent of the short-distance dynamics. In particular, we have
discussed universal expressions for three-body observables and 
universal scaling functions relating various few-body observables
for both atomic and nuclear systems. A more detailed account including
the effects of deeply-bound two-body states can
be found in Ref.~\cite{BrH04}.

The success of the pionless EFT demonstrates that physical
QCD is close to an infrared limit cycle. We have conjectured, that
QCD could be tuned to the critical trajectory for the limit cycle
by adjusting the up and down quark masses. The limit cycle would them
be manifest in the Efimov effect for the triton \cite{Braaten:2003eu}.
It may be possible to demonstrate the existence of this 
infrared RG limit cycle in QCD using Lattice QCD and EFT.

In two spatial dimensions, the three-body parameter $\Lambda_*$ does
not enter at leading order in the expansion in $l/|a|$ and
$N$-body binding energies only depend on $B_2$. The asymptotic freedom
of non-relativistic bosons with attractive interactions in 2D leads
to remarkable universal properties of $N$-body droplets, such as
the exponential behavior of binding energies and droplet sizes in
Eq.~(\ref{RNratio}) \cite{Hammer:2004as}. 
 
The three-body effects discussed here will also
become relevant in Fermi systems with three or more spin states.
Future challenges include universality in the $N$-body problem for 
$N \geq 4$, effective range corrections, and a large scattering length
in higher partial waves. 
A large P-wave scattering length, for example, appears in nuclear Halo systems 
such as $^6$He \cite{Bertulani:2002sz,Bedaque:2003wa}.


\begin{theacknowledgments}
This work was done in collaboration with E.~Braaten, U.-G.~Mei\ss ner,
L.~Platter, and D.T.~Son. It was supported in part by the US Department of
Energy under grant DE-FG02-00ER41132. Local support from the organizers
of the workshop on \lq\lq Nuclei and Mesoscopic Physics'' at NSCL 
at Michigan State University is gratefully acknowledged.
\end{theacknowledgments}


\bibliographystyle{aipproc}   

\begin{thebibliography}{99}

\bibitem{Kaplan:1995uv}
See for example:
D.B.~Kaplan,
arXiv:nucl-th/9506035;
G.P.~Lepage,
arXiv:nucl-th/9706029.

\bibitem{Efi71}
 V.N.~Efimov, Sov.\ J.\ Nucl.\ Phys.\ {\bf 12}, 589 (1971).

\bibitem{Efi79}
 V.N.~Efimov, Sov.\ J.\ Nucl.\ Phys.\ {\bf 29}, 546 (1979).

\bibitem{BrH04}
E.~Braaten and H.-W.~Hammer, arXiv:cond-mat/0410417.

\bibitem{BHvK99}
P.F.~Bedaque, H.-W.~Hammer, and U.~van Kolck,
Phys.\ Rev.\ Lett.\ {\bf 82}, 463 (1999)
[arXiv:nucl-th/9809025].

\bibitem{BHvK99b}
P.F.~Bedaque, H.-W.~Hammer, and U.~van Kolck,
Nucl.\ Phys.\ A\ {\bf 646}, 444 (1999)
[arXiv:nucl-th/9811046].

\bibitem{Sor97}
D.~Sornette,
Phys.\ Rep.\ {\bf 297}, 239 (1998) 
[arXiv:cond-mat/9707012].

\bibitem{Wil71}
K.G.~Wilson, Phys.\ Rev.\ D {\bf 3}, 1832 (1971).

\bibitem{Glazek:2002hq}
S.~D.~Glazek and K.~G.~Wilson,
Phys.\ Rev.\ Lett.\  {\bf 89}, 230401 (2002)
[arXiv:hep-th/0203088].

\bibitem{Glazek:2004}
S.~D.~Glazek and K.~G.~Wilson,
Phys.\ Rev.\ B {\bf 69}, 094304 (2004).

\bibitem{LeClair:2002ux}
A.~LeClair, J.~M.~Roman, and G.~Sierra,
Phys.\ Rev.\ B {\bf 69}, 20505 (2004)
[arXiv:cond-mat/0211338].

\bibitem{Leclair:2003xj}
A.~Leclair, J.~M.~Roman, and G.~Sierra,
Nucl.\ Phys.\ B {\bf 675}, 584 (2003)
[arXiv:hep-th/0301042].

\bibitem{BH02}
E.~Braaten and H.-W.~Hammer,
Phys.\ Rev.\ A\ {\bf 67}, 042706 (2003)
[arXiv:cond-mat/0203421].

\bibitem{BH03}
E.~Braaten and H.-W.~Hammer,
Phys.\ Rev.\ A\ {\bf 70}, 042706 (2004)
[arXiv:cond-mat/0303249].

\bibitem{FTD99}T.~Frederico, L.~Tomio, A.~Delfino, and A.E.A.~Amorim,
 Phys.\ Rev.\ A\ {\bf 60}, R9 (1999).

\bibitem{MSSK01}A.K.~Motovilov, W.~Sandhas, S.A.~Sofianos, 
   and E.A.~Kolganova, Eur.\ Phys.\ J.\ D\ {\bf 13}, 33 (2001).

\bibitem{Phillips68}
A.C.~Phillips, 
 Nucl.\ Phys.\ A {\bf 107}, 209 (1968).

\bibitem{Platter:2004qn}
L.~Platter, H.-W.~Hammer, and U.-G.~Mei\ss ner,
Phys.\ Rev.\ A\ {\bf 70}, 052101 (2004)
[arXiv:cond-mat/0404313].

\bibitem{Blume:2000}
D.~Blume and C.H.~Greene, J.\ Chem.\ Phys.\ {\bf 112}, 8053 (2000).

\bibitem{Tjo75}
J.A.~Tjon, 
Phys.\ Lett.\ B {\bf 56}, 217 (1975).

\bibitem{Platter:tjon}
L.~Platter, H.-W.~Hammer, and U.-G.~Mei\ss ner,
Phys.\ Lett.\ B\ (in press)
[arXiv:nucl-th/0409040].

\bibitem{EfiTk88}
V.\ Efimov and E.G.\ Tkachenko,
Sov.\ J.\ Nucl.\ Phys. {\bf 47}, 17 (1988).

\bibitem{Bedaque:1999ve}
P.F.~Bedaque, H.-W.~Hammer, and U.~van Kolck,
	Nucl.\ Phys.\ A {\bf 676}, 357 (2000)
[arXiv:nucl-th/9906032].

\bibitem{Efi81}
V.~Efimov, Nucl.\ Phys.\ A {\bf 362}, 45 (1981).

\bibitem{Bedaque:2002yg}
P.F.~Bedaque, G.~Rupak, H.W.~Griesshammer, and H.-W.~Hammer,
Nucl.\ Phys.\ A {\bf 714}, 589 (2003)
[arXiv:nucl-th/0207034],
and references therein.


\bibitem{Nogga:2000uu}
A.~Nogga, H.~Kamada, and W.~Gl{\"o}ckle,
Phys.\ Rev.\ Lett.\  {\bf 85}, 944 (2000)
[arXiv:nucl-th/0004023].

\bibitem{Epelbaum:2000mx}
E.~Epelbaum, H.~Kamada, A.~Nogga, H.~Witala, W.~Gl\"ockle,
 and U.-G.~Mei\ss ner,
Phys.\ Rev.\ Lett.\  {\bf 86}, 4787 (2001)
[arXiv:nucl-th/0007057].

\bibitem{Epelbaum:2002vt}
E.~Epelbaum, A.~Nogga, W.~Gl\"ockle, H.~Kamada, U.-G.~Mei\ss ner,
 and H.~Witala,
Phys.\ Rev.\ C {\bf 66}, 064001 (2002)
[arXiv:nucl-th/0208023].

\bibitem{Nogga:2004ab}
A.~Nogga, S.~K.~Bogner, and A.~Schwenk,
Phys.\ Rev.\ C {\bf 70}, 061002 (2004)
[arXiv:nucl-th/0405016].

\bibitem{Beane:2001bc}
S.~R.~Beane, P.F.~Bedaque, M.J.~Savage, and U.~van Kolck,
Nucl.\ Phys.\ A {\bf 700}, 377 (2002)
[arXiv:nucl-th/0104030].

\bibitem{Beane:2002vs}
S.~R.~Beane and M.~J.~Savage,
Nucl.\ Phys.\ A {\bf 713}, 148 (2003)
[arXiv:nucl-th/0206113].

\bibitem{Beane:2002xf}
S.~R.~Beane and M.~J.~Savage,
Nucl.\ Phys.\ A {\bf 717}, 91 (2003)
[arXiv:nucl-th/0208021];

\bibitem{Epelbaum:2002gb}
E.~Epelbaum, U.-G.~Mei{\ss}ner, and W.~Gl{\"o}ckle,
	Nucl.\ Phys.\ A {\bf 714}, 535 (2003)
[arXiv:nucl-th/0207089].

\bibitem{Braaten:2003eu}
E.~Braaten and H.~W.~Hammer,
Phys.\ Rev.\ Lett.\  {\bf 91}, 102002 (2003)
[arXiv:nucl-th/0303038].

\bibitem{Wilson:2004de}
K.~G.~Wilson,
arXiv:hep-lat/0412043.

\bibitem{Hammer:2004as}
H.-W.~Hammer and  D.T.~Son,
Phys.\ Rev.\ Lett.\ {\bf 93}, 250408 (2004)
[arXiv:cond-mat/0405206].

\bibitem{Gross:1973id}
D.\,J.~Gross and F.~Wilczek,
Phys.\ Rev.\ Lett.\  {\bf 30}, 1343 (1973).
 
\bibitem{Politzer:fx}
H.\,D.~Politzer,
Phys.\ Rev.\ Lett.\  {\bf 30}, 1346 (1973).

\bibitem{Bruch}
L.W.~Bruch, and J.A.~Tjon, Phys.\ Rev.\ A\ {\bf 19}, 425 (1979).

\bibitem{Nielsen}
E.~Nielsen, D.V.~Fedorov, and A.S.~Jensen, Few-Body Syst.\ {\bf 27},
15 (1999).

\bibitem{Platter:2004ns}
L.~Platter, H.-W.~Hammer, and U.-G.~Mei\ss ner,
Few-Body Syst.\ {\bf 35}, 169 (2004) 
[arXiv:cond-mat/0405660].

\bibitem{Bertulani:2002sz}
C.A.~Bertulani, H.-W.~Hammer, and U.~van Kolck,
Nucl.\ Phys.\ A\ {\bf 712}, 37 (2002)
[arXiv:nucl-th/0205063].

\bibitem{Bedaque:2003wa}
P.~F.~Bedaque, H.-W.~Hammer, and U.~van Kolck,
Phys.\ Lett.\ B {\bf 569}, 159 (2003)
[arXiv:nucl-th/0304007].

\end{thebibliography}

\end{document}